\documentclass[%
 reprint,
superscriptaddress,
nofootinbib,
 amsmath,amssymb,
 aps,
]{revtex4-2}

\usepackage{comment}
\usepackage{graphicx}
\usepackage{dcolumn}
\usepackage{bm}
\usepackage{hyperref}
\usepackage{amsmath}
\usepackage{braket}
\usepackage{xcolor}
\usepackage{natbib}
\usepackage{placeins}

\usepackage{braket}

\begin{document}

\title{Self-stabilized high-dimensional quantum key distribution on a metropolitan free-space link}

\author{Karolina Dziwulska}
\email{karolina.dziwulska@iof.fraunhofer.de}
 \altaffiliation[]{Authors contributed equally}
\affiliation{Fraunhofer Institute for Applied Optics and Precision Engineering, Albert-Einstein-Strasse 7, Jena 07745, Germany
}%
\author{Christopher Spiess}%
\email{christopher.spiess@iof.fraunhofer.de}

 \altaffiliation[]{Authors contributed equally}
\affiliation{Fraunhofer Institute for Applied Optics and Precision Engineering, Albert-Einstein-Strasse 7, Jena 07745, Germany
}%
\author{Sarika Mishra}%
\affiliation{Fraunhofer Institute for Applied Optics and Precision Engineering, Albert-Einstein-Strasse 7, Jena 07745, Germany
}%

\author{Markus Leipe}%
\affiliation{Fraunhofer Institute for Applied Optics and Precision Engineering, Albert-Einstein-Strasse 7, Jena 07745, Germany
}
  \affiliation{Friedrich Schiller University, Institute of Applied Physics, Abbe Center of Photonics, Albert-Einstein-Strasse 15, Jena 07745, Germany
  }

\author{Yugant Hadiyal}%
\affiliation{Fraunhofer Institute for Applied Optics and Precision Engineering, Albert-Einstein-Strasse 7, Jena 07745, Germany
}%

\author{Fabian Steinlechner}%
\email{fabian.steinlechner@uni-jena.de}
\affiliation{Fraunhofer Institute for Applied Optics and Precision Engineering, Albert-Einstein-Strasse 7, Jena 07745, Germany}
  \affiliation{Friedrich Schiller University, Institute of Applied Physics, Abbe Center of Photonics, Albert-Einstein-Strasse 15, Jena 07745, Germany}

\date{\today}

\begin{abstract}
Quantum communication technologies capable of operating reliably across heterogeneous optical channels are essential for scalable metropolitan quantum networks. Here we demonstrate high-dimensional time-bin-encoded quantum key distribution over a hybrid metropolitan link comprising 1.7 km free-space transmission and 685 m of optical fiber. Operating at a clock rate of 500\,MHz in the C-band, we implement both 2- and 4-dimensional protocols, and obtain estimated secure finite-key rates of ($95\pm28$)\,kbit/s for 4D at ($25.0\pm2.0$)\,dB loss and ($59\pm27$)\,kbit/s for 2D at ($23.5\pm2.3$)\,dB loss.  Crucially, we achieve continuous operation over 48\,h in a fully self-referenced architecture: initial synchronization, interferometric phase stabilization, and long-term drift compensation are performed exclusively using the detected quantum signals, without auxiliary optical reference channels. Our results thus establish a practical and versatile platform for hybrid free-space-to-fiber quantum communication and show that the encoding dimensionality can be adapted to the optimal operating regime of realistic metropolitan channels, providing a pathway toward efficient, autonomous and deployable quantum network nodes.
\end{abstract}

\maketitle

\section{\label{sec:intro}Introduction}
Quantum key distribution (QKD) enables the distribution of encryption keys based on the transmission of quantum states and serves as a central application and benchmark for quantum networks \cite{Zurek1982,BB84,Ekert.1991}. As QKD systems move beyond point-to-point demonstrations toward network deployment, they must operate across heterogeneous optical channels, including metropolitan fiber and free-space links \cite{joshi2020trusted, Krzic.2023,free-space_2024}, long-haul backbone connections \cite{long-haul}, satellite links \cite{li_microsatellite-based_2025}, and hybrid combinations thereof \cite{goy2025ad}. These channels differ substantially in their transmission characteristics, imposing distinct requirements and implementation challenges for QKD transceiver technologies. 
\newline

Ground-to-ground free-space optical (FSO) links are advantageous in scenarios requiring mobility, flexibility, and rapid deployment \cite{Krzic.2023}, while space-to-ground satellite-based approaches enable the distribution of quantum states over global distances \cite{staellite_qkd_review}. A central challenge in all free-space settings arises from the strong variability of environmental conditions with location, time of day, and season, resulting in solar background noise, scintillation, and turbulence-induced wavefront distortions. Practical quantum communication systems must therefore accommodate a wide range of channel conditions, including fluctuating loss and noise levels as well as detector saturation effects, while maintaining stable operation and useful key rates over extended timescales.
\newline

The overall system complexity further depends on the degree of freedom (DOF) used for encoding, such as spatial \cite{PhysRevLett.113.060503,WangLiuLiZhaoDuZhu+2022+645+680}, temporal \cite{cocchi2025timebinencodingquantumkey}, or frequency DOF \cite{frequency_qkd_2025}. Among these, encoding in the temporal DOF is particularly attractive due to its intrinsic compatibility with existing telecommunications infrastructure and its applicability to both optical fiber and free-space channels \cite{cocchi2025timebinencodingquantumkey}. In this approach, quantum information is encoded using discrete time bins, whereby the encoding alphabet is determined by the number of distinguishable time bins. Increasing the alphabet beyond binary corresponds to high-dimensional (HD) quantum state encoding, which can, in principle, increase the information capacity to $\log_2(d)$ bits per detected photon \cite{Cerf.2002,Sheridan.2010,Brougham2013SecurityOH,PhysRevA.87.062322,Zhong.2015,Brougham.2016,Islam.2017,Islam.2019}, enhance robustness to noise \cite{Cerf.2002,Sheridan.2010}, and mitigate detector dead-time and bandwidth limitations in high-count-rate regimes \cite{Islam.2017}. High-dimensional QKD (HD-QKD) has been extensively investigated in laboratory experiments and fiber-based systems \cite{Islam.2017,Islam.2019,Zahidy.2024,Ogrodnik:25,Yu.2025,Efficient,Zhong_2015}.
\newline

\begin{figure*}[htb]
	\centering
	\includegraphics[width=174mm]{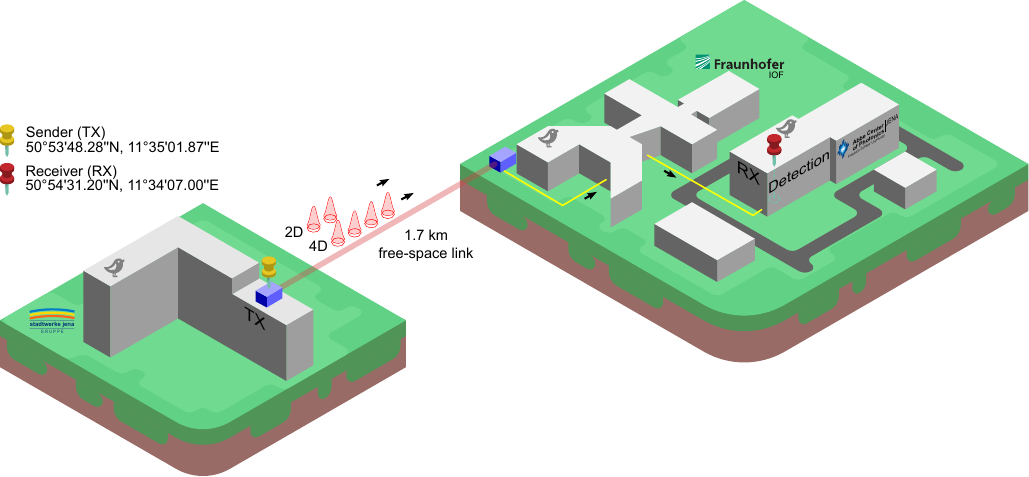}
	\caption[]{\textbf{High-dimensional quantum key distribution on the campus of Fraunhofer IOF and Abbe Center of Photonics.} The weak coherent source is placed in a mobile container on the rooftop of a local provider of energy and water (Stadtwerke Jena Gruppe). 2D and 4D time-bin states are transferred over a 1.7\,km free-space link to another mobile container (QuBUS) located at Fraunhofer IOF, and guided to the nearby Abbe Center of Photonics via fiber and measured with interferometers and nanowire single-photon detectors.}
	\label{fig:1}
\end{figure*}

\begin{table*}
\begin{center}
\begin{tabular}{|c c c c c c c |} 
 \hline
 Year & Reference & DOF & Protocol & Scenario & Loss budget (dB) & Secure key rate \\
 \hline
 2017 & Steinlechner et al. \cite{Steinlechner.2017} & Time-phase-polarization & Entanglement & Free-space & 7.5 & -$^{\ast}$ \\ 

 2017 & Sit et al. \cite{Sit.2017} & OAM & Entanglement & Free-space & 13 & -$^{\dagger}$\\

 2019 & Lee et al. \cite{Lee.2019} & Time-phase & Weak coherent states & Fiber & 12.7 & 1.2 Mbit/s$^{\ddagger}$\\

  2023 & Bulla et al.\cite{Bulla.2023}  & Time-phase & Entanglement & Free-space & 25 & 150 bit/s$^{\S}$ \\
 
  2023 & Liu et al. \cite{Liu_2024} & Time-energy & Entanglement & Fiber & 51 & 0.06 bit/s$^{\ddagger}$\\

  2024 & Zahidy et al. \cite{Zahidy.2024} & Time-path & Weak coherent states & Fiber & 22 & 51.5 kbit/s$^{\ddagger}$\\

  2025 & This work & Time-phase & Weak coherent states & Free-space & 25 & 95 kbit/s$^{\ddagger}$ \\
 \hline
\end{tabular}
\caption{Demonstrations of deployed high-dimensional QKD outside common lab infrastructure
with different degree-of-freedoms (DOF). $^{\ast}$ Estimated sifted key rate of 10 kbit/s,
$^{\dagger}$ 0.43 bit per sifted photon, $^{\ddagger}$ Finite key analysis,
$^{\S}$ Asymptotic key rate.}
\label{tab:results_table}
\end{center}
\end{table*}

In contrast, free-space implementations have so far relied predominantly on two-dimensional polarization or time-bin protocols \cite{cocchi2025timebinencodingquantumkey,Jin.2019,integratedqkd}. Proof-of-principle experiments have explored high-dimensional encoding in spatial modes using heralded single photons \cite{Sit.2017}, demonstrating the feasibility of transmitting high-dimensional quantum states over short terrestrial free-space links, but the approach cannot be scaled to long distance, in particular to satellite links where mode-dependent diffraction loss poses a fundamental obstacle. HD-QKD over free-space links has otherwise been realized primarily in entanglement-based systems employing time-bin and polarization hyperentanglement at wavelengths near 800\,nm \cite{Steinlechner.2017,Bulla.2023,Bulla.2023_pra}, achieving asymptotic key rates of up to 150\,bit/s at channel losses around 25\,dB without fiber coupling \cite{Bulla.2023,Bulla.2023_pra} (summary in Table \ref{tab:results_table}). While these results constitute important proof-of-principle demonstrations, several critical building blocks required for deployment in realistic quantum networks, including efficient fiber coupling, operation at telecom wavelengths, and autonomous long-term operation under daylight conditions have not yet been demonstrated within a single integrated system.
\newline 

In this work, we validate HD-QKD with time-bin encoding under real-world link conditions using a deployed metropolitan testbed that integrates a 1.7\,km free-space link \cite{10.1117/12.2599163, Ji:25} and a 785\,m fiber channel (Fig. 1). Operating at a system clock rate of 500\,MHz, we achieve an estimated secure key rate of $(95 \pm 28)$\,kbit/s at $(25.0 \pm 2.0)$\,dB loss using 4D-encoding \cite{Efficient} and $(59 \pm 27)$\,kbit/s at $(23.5 \pm 2.3)$\,dB loss using 2D-encoding \cite{Boaron.2018}, considering finite-size key effects \cite{1decoy,origfinitekey,wiesemann2024consolidated,20174dIslam} and incorporating decoy states \cite{Hwang_2003_decoy, Wang_2005_decoy, 2005decoy, 1decoy}. The system was operated continuously for up to 48 hours during day and night-time conditions, corresponding to the generation of 4.5\,Gbit of key material. Stable operation on this timescale was enabled by QuantumHarmony, a reference-free scheme for timing synchronization and interferometric phase stabilization based solely on quantum-signal-based feedback \cite{PhysRevApplied.19.054082,Spiess_2024,Calderaro.2020}. The impact of noise sources is reduced by tight spectral, temporal, and spatial filters \cite{Krzic.2023,Krzic:23}. Together, these results demonstrate the practical feasibility of HD-QKD over hybrid free-space and fiber links and establish encoding dimensionality of time-bin implementations as a configurable parameter for adapting QKD systems to realistic metropolitan channel conditions.
\newline

\section{\label{sec:res}Results}

\begin{figure*}[!htb]
	\centering
	\includegraphics[width=142.105mm]{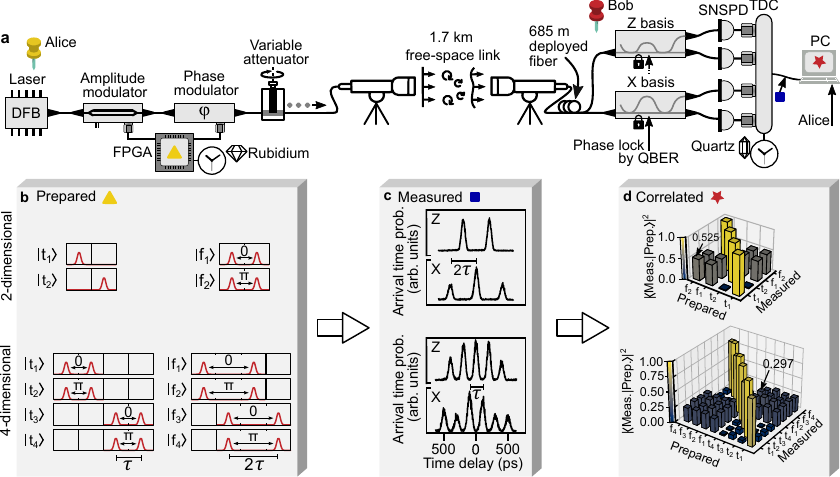}
	\caption[]{\textbf{Experimental setup for higher dimensional encoding.} \textbf{a}, The setup consists of a distributed feedback laser (DFB) and electro-optical modulators to prepare the time-bin states. \textbf{b}, The state sequence is loaded and output by a FPGA that is locked to an external rubidium clock. The photons are sent through telescopes to the remote receiver. The free-space link infrastructure includes tip-tilt wavefront correction through beacon lasers. Noise is filtered by coupling the 1550\,nm light into a single-mode fiber that acts as a spatial filter. The receiver incorporates two interferometers that are phase-locked using the quantum bit error rate obtained in the parameter estimation step as the error signal. \textbf{c}, The arrival time is measured through nanowire single-photon detectors (SNSPD) and time-to-digital converters (TDC). \textbf{d}, Probability of detection when each input state is detected in each interferometer, as characterized in laboratory conditions without the free-space channel. }
	\label{fig:2}
\end{figure*}

\textbf{QKD system and calibration.} The QKD system consists of a sender (Alice), who prepares weak coherent pulses with quantum states encoded in the time-bin DOF, and a receiver (Bob), who decodes the states using unbalanced interferometers and detects them with superconducting nanowire single-photon detectors (Fig.~\ref{fig:2}\textbf{a}). The state sequence with a length of 1000 symbols is encoded in a field programmable gate array (FPGA) in an infinite loop. The sequence is long enough to provide reliable statistics for the security parameters, actively phase-stabilize the interferometer, and run our synchronization protocol without losing generalization. 
\newline

In the two-dimensional (2D) time-bin encoding, the Z basis states are directly encoded in the early $\ket{\text{E}}$, late $\ket{\text{L}}$ time bins, while in the X basis states are encoded in the relative phase between time bins, yielding the \textit{phase} states ``$0 \pi$'' $\ket{\text{+}}$, ``$1 \pi$'' $\ket{-}$. This type of encoding is the first experimental demonstration reported here. 
\newline

In four-dimensional (4D) time-bin encoding, a similar encoding strategy can be implemented \cite{Islam.2017}, with the Z basis consisting of the time states $\ket{t_n}$ ($n = 0,1,2,3$) and with the X basis consisting of the phase states $\ket{f_n}$ ($n = 0,1,2,3$). In this approach, the X basis measurement requires a cascade of interferometers, which increases complexity and reduces the practicality of high-dimensional encoding \cite{Islam.2019}. To address this issue, Vagniluca et al. \cite{Efficient} proposed a strategy that reduces the required interferometric depth by mixing time and phase states (Fig. \ref{fig:2}\textbf{b}). This encoding scheme is used in the second experimental demonstration presented in this work.   
\newline

During preliminary calibration in the laboratory, arrival time distributions corresponding to different basis states (Fig. \ref{fig:2}\textbf{c}) are identified. The temporal separation between time bins is $2\tau = 400 \,$ps and $1\tau = 200\,$ps for 2D and 4D, respectively. This exceeds the average root mean square (RMS) timing jitter of $\sigma = (28.9 \pm 2.4)\,$ps of the detectors and time-tagging system used, and thus allows us to further filter the corresponding states. Arrival-time filtering to 100\,ps ($\approx 1.6\sigma$) suppresses cross-talk of adjacent time bins at the expense of approximately 11\,\% of signal loss. The filtered arrival times are correlated and compared with the prepared states, yielding the probability distribution shown in Fig. \ref{fig:2}\textbf{d}. The states in the X basis should be mutually unbiased with respect to the states in the Z basis, with the probability depending on the dimensionality $d$, $|\braket{t_n|f_n}|^2=1/d$. The highest correlation values occur for states that are prepared and also measured in the same basis (diagonal elements). From this calibration, we determine an overlap parameter $c = -\log_2 \text{max}_{i,j}|\braket{t_i|f_j}|^2$, yielding 0.93 bits in the two-dimensional case and 1.75 bits in four dimensions (Fig. \ref{fig:2}\textbf{d}). These values are reduced with respect to the ideal value of $\log_2(d)$ due to non-ideal state preparation. The state preparation quality of the source has been approximated with the c parameter according to \cite{20174dIslam}. Rigorous implementation of this parameter into the security proof remains outside of the scope of this work. Extending the security analysis to relax this assumption is an important direction for future work and has only recently begun to be explored such as in references \cite{tupkary2025qkdsecurityproofsdecoystate,kanitschar2025composablefinitesizesecurityhighdimensional, PhysRevApplied.23.044011, Ogrodnik:25}.
\newline

\textbf{Security analysis.} 
The security analysis in this work is based on the smooth min- and max-entropy \cite{rennerthesis, 5208530}, specifically the generalization of the uncertainty relation for smooth entropies (EUR) \cite{smoothentropies}. We use the 1-decoy state finite-key analysis presented in \cite{origfinitekey, 1decoy}, which was modified for the 4D QKD in \cite{20174dIslam}. The finite-key protocol with decoy states \cite{Hwang_2003_decoy, 2005decoy, Wang_2005_decoy, PhysRevA.98.012330, wiesemann2024consolidated} considers the statistical fluctuations of the link parameters and is resilient to the photon number splitting attack \cite{GLLP}. The secret key length $l$ is calculated as given by equation \ref{eq:security_proof}. Here, $d$ is the dimensionality, $s_{Z,0}^{l}$ is the lower bound on the zero-photon events, and $s_{Z,1}^{l}$ is the lower bound on the single-photon events in the Z-basis and $c$ is the overlap parameter (obtained from prior in-lab calibration). The function $H_d$ denotes the binary entropy for dimension $d$, and $\phi_{Z}$ is the phase error rate as inferred from the X-basis measurements. The parameters $a$ and $b$ denote the decoy state parameters (for 1-decoy state protocol $a=6$, $b=19$). $\Delta_{EC}$ upper-bounds the information leakage during error correction. Finally, $\epsilon_{sec}$ and $\epsilon_{cor}$ are the secrecy and correctness parameters, where the secrecy parameter upper bounds the probability of distinguishability of the real state from its ideal instance, and the correctness parameter upper-bounds the probability of mismatch between Alice's and Bob's keys. Throughout the experiment, the mean photon number is tracked by tapping optical power at the sender prior to attenuation. This leads to live-adjustments of the mean photon values used in the security proof and in consequence of the parameters $s_{Z,0}^{l}$ and $s_{Z,1}^{l}$. This way Alice and Bob remain aware of changes in their system parameters during protocol implementation. 
\newline

\begin{widetext}
\begin{equation}
    l\leq \log_2(d)\cdot s_{Z,0}^{l} +  s_{Z,1}^{l}\cdot\left[c - H_d(\phi^{\text{tol.}}_{Z})\right] - \Delta_{EC} - a \cdot \log_2\left(\frac{b}{\epsilon_{sec}}\right) - \log_2\left(\frac{2}{\epsilon_{cor}}\right) \label{eq:security_proof}
\end{equation}
\end{widetext}

\begin{figure*}[htb]
	\centering
	\includegraphics[width=167mm]{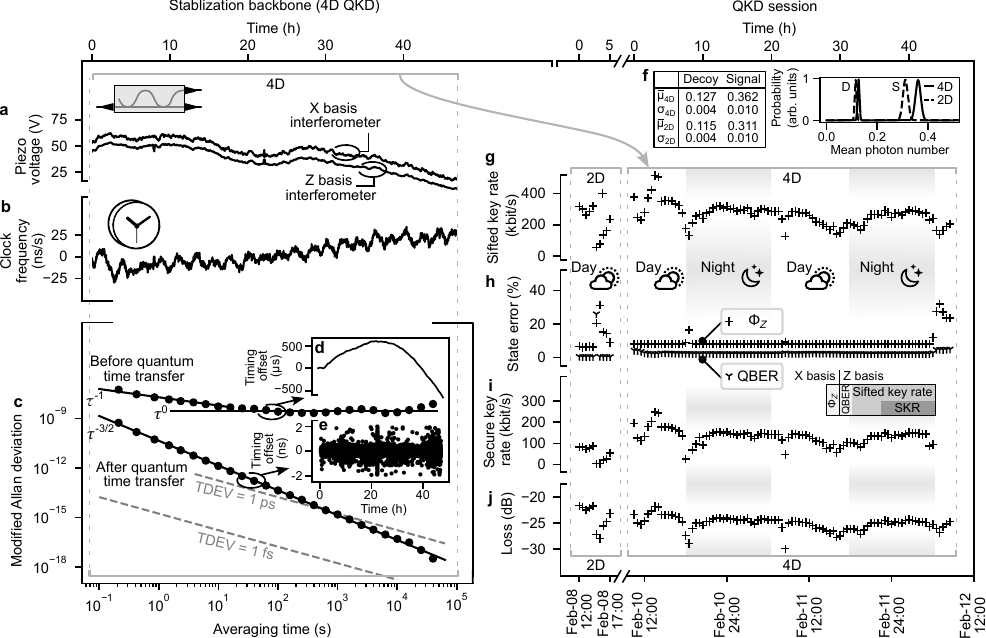}
	\caption[]{\textbf{Results with 2-dimensional and 4-dimensional encoding on the 1.7 km free-space link.} \textbf{a}, Piezo control voltage that is sent to the interferometer for stabilization. \textbf{b}, Difference in the relative frequency of the rubidium oscillator at sender and quartz oscillator at the receiver. \textbf{c}, Timing stability before and after quantum time transfer. The noise figures improve from flicker phase modulation noise ($\tau^{-1}$) and flicker frequency modulation noise ($\tau^{0}$) to only white phase modulation noise  ($\tau^{-3/2}$). \textbf{d}, The timing offset varies by up to 1\,ms before stabilization. \textbf{e}, After stabilization, the fluctuation of the timing offset reduce substantially with standard deviation of 58\,ps. \textbf{f}, Distribution of mean photon number over the full measurement time. \textbf{g}, Sifted key rate extracted from both Alice and Bob choosing the Z basis. \textbf{h}, State error rate in Z basis estimated from X basis $\Phi_Z$ and quantum bit error rate (QBER) estimated from a 10\,\% subsample of states in Z basis. \textbf{i}, The finite secure-key rate. Data over 10\,seconds is acquired and considered with finite-size key effects. Every data point represents the mean key rate over 30\,min. All information in X basis is published and used for estimating the error $\Phi_Z$, as well as to stabilize the X-basis interferometer. In the Z basis, 10\,\% of the detections are used to estimate the QBER for error correction and to stabilize the Z-basis interferometer. \textbf{j}, End-to-end estimation of loss on the link, based on the single-photon count rate. The subplots \textbf{a,b,c,d,e} refer to the 4-dimensional protocol. All data was recorded on 8$^{\text{th}}$-12$^{\text{th}}$ February 2024 (see Methods \ref{sec:Environmental and turbulence measurements}, Fig. \ref{fig:fig_8_envirnmental_measurements} for weather and turbulence data).}
	\label{fig:experiment}
\end{figure*}

\textbf{Stabilization backbone.} The initial global timing offset is calculated by a cross-correlation of the prepared and received time-bin states. To do so, the state information of a complete 100-ms dataset is published and correlated once to obtain high signal-to-noise ratios and high statistical confidence. Note that this dataset is not used as key material but rather as a synchronization pattern \cite{Spiess_2024,Calderaro.2020}. After initialization and during the session, the arrival time of photons is used \textit{locally} for live-adjustment of the clock frequency and timing offset between quartz and rubidium oscillator. When the QBER appears to be 50\,\%, it is likely that sent and received sequence are mismatched by at least 1 clock cycle. In this case, Bob locally introduces sufficient timing offset to identify the mismatch. Synchronization is successful and continuously confirmed by low QBERs in parameter estimation.    
\newline

To achieve long-term phase stability, we implement closed-loop locking of interferometers using the QBER as the error signal. For this purpose, 10\,\% of the sifted bits in the Z-Basis are publicly revealed to estimate the QBER, while the remaining 90\,\% proceed in the pipeline for secure key generation. Although this reduces the effective duty cycle for key generation by 10\,\%, the disclosed measurement results still fulfill multiple roles. They provide an instantaneous estimate of the number of parity checks required during error correction, contribute to upper-bounding zero-photon events required in the 1-decoy state protocol and are utilized for monitoring the phase drift in the Z-basis interferometer in 4D encoding. In the X-basis, all sifted bits are used for parameter estimation as well as for locking the interferometer there \cite{Spiess:23}. 
\newline

We yield high temporal and phase stability on the 1.7\,km free-space link through live tuning of the piezo voltage towards the interferometer and by live adjustment of the timing offset through an algorithm customized to high-dimensional systems. The phase drifts within the interferometers are caused by slow changes of the laboratory temperature (Fig. \ref{fig:experiment}\textbf{a}). The clock frequency is adjusted live by tracking the timing offset in a peak-finding algorithm (Fig. \ref{fig:experiment}\textbf{b}) and leads to a substantial reduction of the timing uncertainty in contrast to before the lock. After applying the algorithm for quantum-based time transfer, presented in detail in Ref. \cite{Spiess_2024}, the variation of the timing offset is dominated by white-phase modulation noise (Fig. \ref{fig:experiment}\textbf{c}). It eliminates any systematic drift of the timing offset from up to 1\,ms over 48\,hours (Fig. \ref{fig:experiment}\textbf{d}) down to a standard deviation (RMS) of 58\,ps (Fig. \ref{fig:experiment}\textbf{e}) and provides an Allan time deviation of 2.52\,ps and 0.89\,ps when averaging over 100\,s and 1000\,s, respectively. The timing jitter is smaller than previously (timing offset ranges from -1.5\,ns to +1.5\,ns (3$\sigma$), RMS $\approx$ 500\,ps (1$\sigma$) \cite{Calderaro.2020}), is similar to (30-50)\,ps in GPS-synchronized systems \cite{Steinlechner.2017,Ecker2021_island} and the long-term stability is better than other demonstrations running on two Rb-vapor cell clocks (Allan time deviation 88\,ps \cite{Lee.2022} and 41\,ps \cite{Quan.2022} for 100\,s integration time). 
\newline
\newline

\textbf{High-dimensional QKD through turbulent atmosphere.} The free-space link is established between two custom-made telescopes located in mobile shipping containers \cite{goy2021high,Goy_2025}. The telescopes are implemented as obscuration-free afocal mirror telescopes with an entrance aperture of 200\,mm and magnification of $20\times$ \cite{10.1117/12.3022982} (see Methods \ref{sec:Optical alignment of tip-tilt system}). The source is deployed in a container located on the roof of the Stadtwerke Jena (Alice), and transmitted to Bob via a 1.7\,km free-space link. The photons are received by the telescope, coupled to a single-mode fiber using active tip-tilt correction, and guided to the receiver station (located in the Abbe Center of Photonics) through 685\,m of fiber. During the operation of the QKD session, the mean photon number is continuously tracked (Fig. \ref{fig:experiment}\textbf{f}). \newline

The mean sifted key rate from both Alice and Bob choosing the Z-basis amounts to ($208\pm80$)\,kbit/s (2D) and ($266\pm35$)\,kbit/s (4D) (Fig. \ref{fig:experiment}\textbf{g}). The QBER in the 2D encoding has a mean value of ($0.5\pm0.37$)\,\%, being significantly lower than ($2.47\pm0.62$)\,\% considering the 4D encoding (Fig. \ref{fig:experiment}\textbf{h}). We attribute this difference to the fact that in 2D no phase states are used for the key, which reduces the complexity. In 4D, both bases are composed of superposition states that needed active phase stabilization. \newline

The mean secure-key rate is ($59\pm27$)\,kbit/s at ($23.5\pm2.3$)\,dB loss and ($95\pm28$)\,kbit/s at ($25.0\pm2.0$)\,dB loss for 2D and 4D, respectively, with optimal simulated error correction with an inefficiency of $f_e = 1.08$ \cite{efficient_EC} as often used in literature for better comparison (Fig. \ref{fig:experiment}\textbf{i,j}). The estimate of the finite secure key rate is extracted from the accumulated sifted bits over 100\,s, generated by 1000 data points that each have an integration time of 100\,ms. The short integration time allows for fast feedback loops for locking the interferometers. \newline

\begin{figure}[htb]
	\centering
	\includegraphics[width=74.419mm]{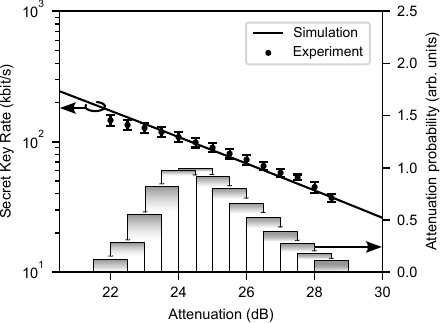}
	\caption[]{\textbf{Variation of the 4D-QKD secure key rate under atmospheric turbulence.} Atmospheric turbulence leads to fluctuations in the channel transmission, enabling the extraction of secure key rates over a broad range of attenuation values. Each recorded data point, acquired within 100 ms integration time, is sorted into 1\,dB-wide attenuation bins according to the instantaneous link transmission, which follows the probability distribution indicated in the bar chart. A finite key analysis is performed for each attenuation bin, introducing a block size of $n_Z = 10^7$. Note that we only consider data points acquired during night-time operation. During daytime operation, fluctuations in background-light-induced noise counts introduced significant bias in channel-loss estimates derived from detected count rates. Simulation parameters: probability of an error = 1\%, probability of a dead count = 4e-07, dead time of the detectors = 5e-08, probability for both Alice and Bob to choose the Z basis = 90\%, probability to choose the signal state = 76\%, signal state mean photon number = 0.37, decoy state mean photon number = 0.13, $\epsilon_{cor} = 10^{-9}$, $\epsilon_{sec} = 10^{-15}$, generation rate 500\,MHz, fidelity of the states $c = 0.297$, error correction inefficiency $f_e$ = 1.08.}
	\label{fig:experiment_attenuation_4D}
\end{figure}

The atmospheric turbulence causes fluctuation of the link transmission (see Methods \ref{sec:Environmental and turbulence measurements}), which allows us to estimate the finite secure-key rate for different loss scenarios. For that, we only take the data points from the night-time measurement to avoid any residual noise which is not covered in our simulation model. The sifted key length and errors are sorted into bins that each correspond to an attenuation range. The values are then summed until a finite block size of $10^7$ is reached and a finite secure key length is calculated. The mean finite secure-key rate for a given attenuation value matches the analytical trendline (Fig. \ref{fig:experiment_attenuation_4D}). 
\newline

\section{Discussion}
The present work establishes autonomous quantum key distribution based on time-bin encoded symbols in 2D and 4D variants over a metropolitan free-space–fiber link and reveals central practical trade-offs between dimensionality and performance.

Quantum-based synchronization in the 4D-QKD time-bin system highlights an important practical trade-off associated with increasing temporal dimensionality under realistic link conditions with drifting reference clocks. In our implementation, the receiver relied on a quartz oscillator with comparatively large frequency drift, while the free-space channel introduced strong turbulence-induced intensity fluctuations. In the 4D protocol, quantum states are distributed over a larger number of temporal slots: 5 in the Z basis and 6 in the X basis - compared to only 2 and 3 slots, respectively, in the 2D implementation. This temporal dilution reduces the effective signal-to-noise ratio per slot, making the extraction of reliable temporal correlations more susceptible to fluctuations in received photon flux.

As a consequence, periods of strong atmospheric turbulence occasionally lead to severely distorted arrival-time probability distributions. During approximately 25\,\% of the measurement time, the observed distribution was distorted to an extent that it prevented the localization of the temporal frame, resulting in a temporal mismatch of 1-3$\tau$ and unsuccessful key extraction for those blocks. Importantly, this behavior does not represent a fundamental limitation of HD encoding, but rather reflects the increased sensitivity of temporally extended alphabets to varying link conditions, intensity fluctuations, and clock instabilities. This challenge is expected to become more pronounced as the temporal dimensionality of encoding increases, highlighting the need for improved postprocessing algorithms or more robust time references when operating time-bin HD-QKD over turbulent free-space links. \newline

Beyond synchronization, increasing dimensionality also impacts the complexity of the required interferometric stabilization. In the present experiments, we reserved 10\,\% of the correlations in the Z-basis for interferometric stabilization, thus reducing the fraction available for secure key generation to 90\,\%. In HD implementations relying on cascaded interferometers and quantum-based stabilization, this overhead would increase further, and will also incur additional loss due to postselection, thereby reducing the secure key fraction. Promising mitigation strategies include post-selection free interferometers \cite{Vedovato.2018,Steinlechner.2017,Carvacho.2015} or dedicated alignment patterns that are dedicated to interferometer stabilization only. Folded interferometers \cite{Ikuta.2022} or path-encoded schemes \cite{Zahidy.2024} are also promising avenues to reduce the number of interferometers needed.
\newline

To compare the performance of the 2D and 4D protocols under identical technological constraints, the relevant limitation is the minimum time-bin duration, predominantly set by timing jitter. For a comparison on equal footing, we also compare the two by taking account for the factor-of-two scaling of the symbol rate associated with the larger alphabet. In the absence of noise and implementation imperfections, one symbol in 4D encoding would ideally provide the capacity of two independent 2D-encoded symbols, as four temporal modes in a 4D-qudit reserve the same time frame as the temporal modes of two qubits.

After rescaling the 4D symbol repetition rate to match the temporal constraints of the 2D implementation, we compare both protocols at identical channel loss while explicitly accounting for the increased symbol duration. Under these conditions, the inferred mean secure key rate of the 4D protocol is 67\,kbit/s, which lies within the statistical confidence interval of the 2D implementation. The reader finds more details on the comparison in Fig. \ref{fig:2D and 4D experiment comparison} (model with identical experimental parameters) and Fig. \ref{fig:skr_model} (model with optimal theoretical parameters) in the Methods \ref{sec:exp_sim_opt}.

The comparable performance for 2D and 4D protocols arises because the system operated well below the saturation regime of the superconducting nanowire detectors (dead time of 50\,ns). Throughout most of the experiment, the single-photon count rate per detector remained below $380\times10^3$\,counts/s (see Table \ref{tab:parameter_qkd_session}). Numerical simulations indicate that for loss regimes below approximately 10\,dB, where detector dead time becomes a performance-limiting factor, 4D-encoding provides a clear advantage in our experimental configuration by mitigating detector saturation effects (Methods \ref{sec:methods_num_optim}). 

However, this equivalence is highly implementation specific (see Fig. \ref{fig:experiment_attenuation}). It nevertheless highlights the potential of four-dimensional encoding as a viable design choice for future QKD protocols, including long-distance scenarios. 
\newline

In summary, our results demonstrate that both two- and four-dimensional time-bin QKD can be operated reliably over metropolitan free-space links when supported by autonomous, quantum-signal-based synchronization and stabilization. At the same time, increasing dimensionality amplifies sensitivity to channel-induced fluctuations and stabilization overhead, confirming that the choice of dimensionality must be carefully optimized with respect to link distance, channel loss, and detector characteristics \cite{Zhong_2015,Lee.2019}. Consistent with previous studies on optimal dimensionality in quantum information tasks \cite{zhu2021high}, we find that the optimal dimensionality of time-bin encoding is a nuanced, system- and link-dependent parameter rather than a universally optimal number. These findings point toward scalable and versatile quantum communication networks in which the encoding dimensionality can be dynamically tailored to channel conditions, enabling efficient and flexible secure key distribution across future network infrastructures.

\FloatBarrier

\begin{acknowledgments}
Christopher Spiess, Markus Leipe, and Fabian Steinlechner are part of the Max Planck School of Photonics supported by BMFTR, Max Planck Society, and Fraunhofer Society. The authors thank the Stadtwerke Jena GmbH for providing access to their infrastructure and hosting the inter-city free-space link. Furthermore, the authors are very thankful for the valuable discussions with Jerome Wiesemann and Jan Krause regarding the security of the protocols. 
\end{acknowledgments}

\section*{Data Availability Statement}
The data that support the findings of this study are available from the corresponding authors upon reasonable request.

\section*{Funding}
This research was conducted within the scope of the project QuNET, funded by the German Federal Ministry of Education and Research (BMBF) in the context of the federal government’s research framework in IT-security “Digital. Secure. Sovereign.”. The authors acknowledge support by the Carl-Zeiss-Stiftung within the Carl-Zeiss-Stiftung Center for Quantum Photonics (CZS QPhoton) under the project ID P2021-00-019.

\begin{figure*}[htb]
	\centering
	\includegraphics[width=171mm]{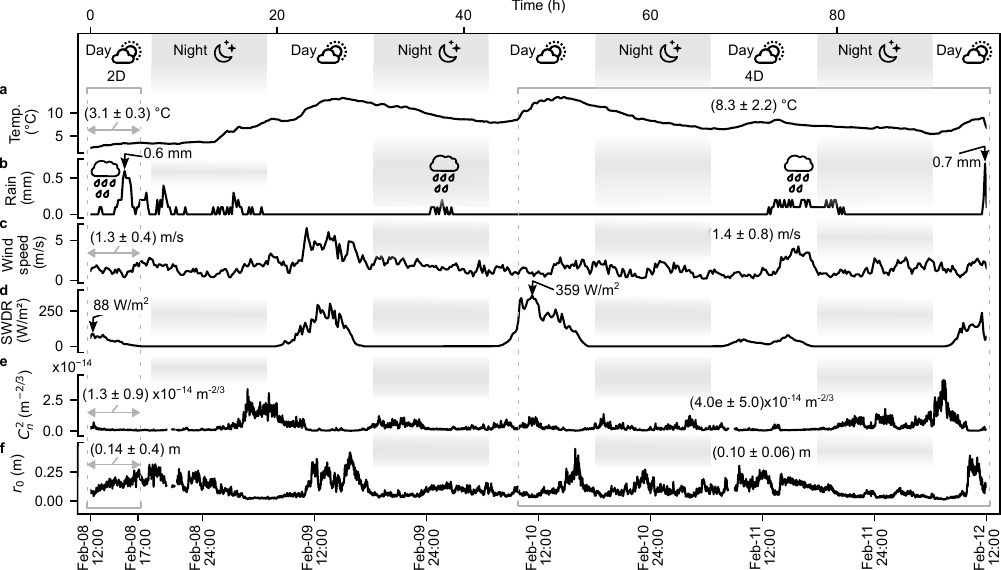}
	\caption[]{\textbf{Environmental and turbulence data during the experiment.} \textbf{a}, The temperature (Temp.) with mean value and standard deviation during the 2D and 4D QKD experiments. \textbf{b}, Rain data with maximum values. \textbf{c}, Wind speed with mean value and standard deviation during the 2D and 4D QKD experiments. \textbf{d}, Surface Shortwave Downward Radiation (SWDR) with maximum values. \textbf{e}, Turbulence parameter $C^2_n$ with mean value and standard deviation during 2D and 4D QKD experiments. \textbf{f}, Fried parameter $r_0$. For reference, our beam diameter is approximately 8\,cm during free-space transmission. The weather data (\textbf{a-d}) was sourced from the weather station Beutenberg, Max-Planck-Insitut für Biogeochemie, Jena. The turbulence data (\textbf{a-d}) has been measured with a scintillometer. All data was recorded on 8$^{\text{th}}$-12$^{\text{th}}$ February 2024.}
	\label{fig:fig_8_envirnmental_measurements}
\end{figure*}

\newpage

\section{\label{sec:con}Methods}

\subsection{Optical alignment of tip-tilt system}\label{sec:Optical alignment of tip-tilt system}
The free-space link was realized using two 200\,mm aperture obscuration-free, off-axis four-mirror telescope transceivers, both equipped with active 4D beam stabilization for atmospheric turbulence correction. Both telescopes have a 20x magnification for a $1/e^2$ beam diameter at the sender aperture of 64 mm. A beacon laser at 1064\,nm was co-propagated alongside the 1550\,nm quantum signal and detected with a position-sensitive device in order to provide the error signal for stabilization. In parallel, the atmospheric turbulence strength (refractive index structure parameter $C_n^2$) along the beam path was continuously measured using scintillometer. We note that a beacon laser is employed exclusively for spatial beam pointing and fiber-coupling stabilization of the free-space link, and does not carry timing or phase information related to the quantum states. 

\subsection{Environmental and turbulence measurements}\label{sec:Environmental and turbulence measurements}
Fig. \ref{fig:fig_8_envirnmental_measurements} summarizes the measurements of turbulence and meteorological  conditions recorded during the 2D and 4D experiments. The turbulence parameters (fried parameter and atmospheric turbulence strength) were obtained using a commercial scintillometer, while the weather data was sourced from the Max Planck Institute for Biogeochemistry weather station in Jena.

Both experiments were conducted under varying meteorological and turbulence conditions, spanning both daytime and nighttime. 
The experiment with the 2D protocol was carried out under conditions of higher Fried parameter and higher humidity than the 4D experiment. The 2D experiment was done with low solar radiation. 
In contrast, the 4D protocol was operated for a longer duration, and subjected to stronger temperature fluctuations, increased sun radiation (ranging from moderate at daytime to none at nighttime), higher wind speed, and enhanced turbulence. The results underline the robustness of the high-dimensional protocol for a broad range of environmental conditions.

\subsection{Experimental parameters, and measurement and simulation comparison}\label{sec:exp_sim_opt}
\begin{figure}[htb]
	\centering
	\includegraphics[width = 80 mm]{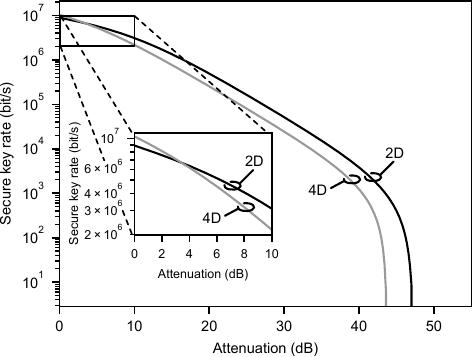}
	\caption[]{Optimization of the secure key rate for 4- and 2- dimensional protocol with the experimental parameters from the measurement. The generation rate is 500 MHz, the dead time is 5e-08 s, the probability of a dark count is 4e-07, error probability is 1\%, and $n_Z$ = $10^7$ for both protocols, state preparation with $c = 1.75$\,bits (2D: $c = 0.93$\,bits). The dots are the mean and max simulation values that correspond to the measured values from the experiment.}
	\label{fig:2D and 4D experiment comparison}
\end{figure}

This section presents the results of the numerical simulation of the link behavior for the parameters and conditions set in the experiment in Fig. \ref{fig:2D and 4D experiment comparison} .

\begin{table*}
\begin{center}
\begin{tabular}{|c c c c c c c c c c c c c|} 
 \hline
 D & $\mu_{1}$ & $\mu_{2}$ & $p_{\mu_{1}}$& Ch1 (kcps) & Ch2 (kcps) & Ch3 (kcps)& Ch4 (kcps) &  $R_{\text{sift}}$ (kbit/s) & QBER (\%) & $\Phi_Z$ (\%) & $R_{\text{sec}}$ & $K_{\text{tot}}$ \\
 \hline\hline
 2 & 0.35 & 0.13 & 0.73 & $325\pm137$ & $325\pm137$ & $60\pm29$ & - & $208\pm80$ & $0.5\pm0.37$ & $3.79\pm1.03$ & 59\,kbit/s & 0.48\,Gbit \\ 
 \hline
 4 & 0.37 & 0.13 & 0.76 & $249\pm97$ & $270\pm101$ & $53\pm24$  & $86\pm32$ & $266\pm35$ & $2.47\pm0.62$ & $7.03\pm4.34$ & 95\,kbit/s & 4.57\,Gbit \\
 \hline
\end{tabular}
\end{center}
\caption{Parameters of the 8-hours and 48-hours QKD session for the protocol with dimensionality $D = 2$ and $D = 4$, respectively. The table lists the single-photon count rate for each channel. In 2D, channel Ch1 represents detections in the Z basis. Channel Ch2 and Ch3 correspond to the X basis. In 4D, the channel Ch1 and Ch2 represent the Z basis and channel Ch3 and Ch4 represent the X basis. Further parameters are: sifted key rate $R_\text{sift}$, QBER, phase error rate $\Phi_Z$, secure key rate $R_\text{sec}$ and total secure bits for the complete QKD session $K_\text{tot}$. }
\label{tab:parameter_qkd_session}
\end{table*}

The dead time of 5e-08 s and the same generation ratio in the experimental scenario give advantage to the 2D protocol for attenuation values over 3,5 dB.  Fig.~\ref{fig:2D and 4D experiment comparison} represents the simulation of the 2D and 4D protocol with the parameters set in the experiment. The simulation shows a higher secret key rate for the 2D protocol for our range of attenuation; however, the difference between the two falls into the tolerance range of the experimental results.

\subsection{Numerical optimization}\label{sec:methods_num_optim}
\begin{figure}[htb]
	\centering
	\includegraphics[width = 86 mm]{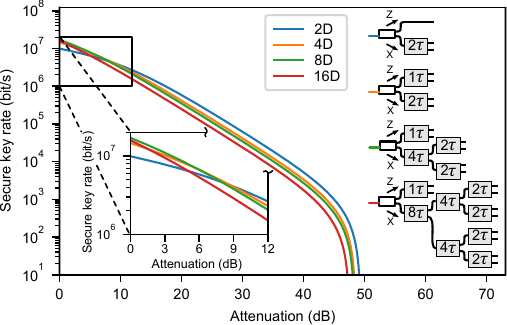}
	\caption[]{Secret key rate optimization for different dimensionalities for $n_Z$ = $10^7$, and detection setup including the interferometric scheme for each dimensionality.}
	\label{fig:skr_model}
\end{figure}

This section relates to the numerical simulation and parameter optimization of the secret key rate for different dimensionalities of the link with equal time-bin duration achieved through generation rate scaling. The results are presented in Fig. \ref{fig:skr_model}.

The parameters used in the experiment were optimized in the finite-key scheme with one decoy state according to \cite{1decoy}, \cite{20174dIslam}, and \cite{Efficient}. The probability of Alice choosing states of the Z basis is 90\,\%. This comes from that fact that we require sufficient number of successful detections in X-basis for stabilization of the interferometer there. Successful means that both Alice and Bob choose the X basis and Bob filtered out detections with the correct arrival times as required in interferometers which do not employ post-selection-free setups. For the 1-decoy state protocol, the mean photon numbers of the signal ($\mu_1$) and decoy ($\mu_2$) states must meet the following condition:
\begin{equation}
 \mu_2<\mu_1<1   
\end{equation}

The simulation in Fig. \textbf{\ref{fig:skr_model}} assumes the setups designs presented in the figure and lossless detectors (the experimental model in Fig. \textbf{\ref{fig:experiment_attenuation_4D}} takes into account the attenuation form the interferometers used in the experiment). The measurement setup requires post-selection of the arrival times as not all photons interfere. In 2D encoding, only 50\% of the photons interfere in the X basis and can be used for eavesdropping analysis. As we have an interferometer in the Z basis in our 4D implementation, only 50\% of qudits proceed with sifting there, which lowers the raw key rate ($n_Z$/time). In our simulation, we consider 3-dB loss of events in the Z basis due to postselection, except in 2D encoding. In the X-basis, we introduce 3\,dB, 3\,dB, 6\,dB and 9\,dB of event loss for 2D, 4D, 8D, and 16D encoding, respectively, due to the chain of interferometers increasing with the dimension. 
The parameters for the optimization were: the probability of a dark count $P_{DC} = 4\cdot10^{-7}$ per clock cycle, the dead time $t_{DT} = 5\cdot10^{-8}$\,s, the error probability 1\%, the secrecy and correctness parameters were $10^{-9}$ and $10^{-15}$, and the block size $n_Z = 10^7$.\\

\begin{figure}[htb]
	\centering
	\includegraphics[]{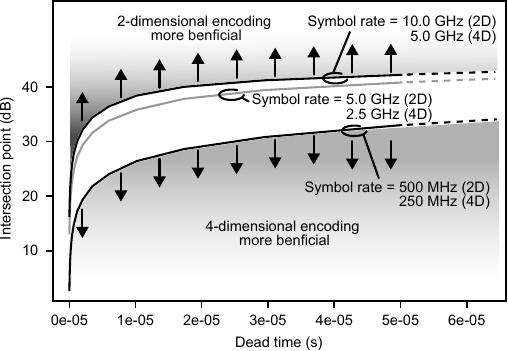}
	\caption[]{Attenuation intersection point (analoguous to Fig. \ref{fig:skr_model}) of 2D and 4D protocols for different dead times of detectors. Link attenuation for which 2D protocol starts to have advantage over 4D protocol for different dead times of the detectors, for $n_Z$ = $10^7$, $P_{DC} = 4\cdot10^{-7}$, $R = \frac{1}d{\tau}$ and error probability of 1\,\%. The trade-off point depends on the system clock rate point. Given a high clock rate, higher-dimensional encoding is more beneficial for a larger attenuation range.}
	\label{fig:experiment_attenuation}
\end{figure}

In the optimization, the dead time of the detectors is simulated according to ref \cite{1decoy}.\\
The generation rate for the simulation was 500 MHz for the 2D protocol and scales as $R = \frac{1}d{\tau}$, where $\tau$ is the duration of the time bin. \\
\\
The dimensionality affects the secret key rate through multiple competing mechanisms. While higher dimensions increase the information content per detected symbol, they also require longer symbol durations and introduce additional losses due to the increased complexity of the interferometric detection scheme. Furthermore, detector dead time and dark counts impact different dimensionalities unequally, depending on the operating regime. As a result, the net performance advantage of higher-dimensional encoding is sensitively dependent on the balance between channel attenuation, detector saturation, and interferometric loss. Fig. \ref{fig:experiment_attenuation} presents the intersection point for 2D and 4D protocol with different detector dead times. It shows the attenuation range where the 4D protocol remains beneficial in relation to the 2D protocol. 
\\
The 2D and 4D protocols constitute a particularly interesting case, as in the absence of implementation losses they are expected to yield the same secret key rate. In practice, however, additional losses introduced by the interferometric detection scheme reduce the performance of the 4D protocol, such that it ceases to be advantageous once the operation moves beyond the detector dead-time-limited regime. Moreover, the secret key rates of the 2D and 4D protocols differ due to an error-correction penalty term that does not scale with the dimensionality, as well as the advantage of higher information capacity is counterbalanced by scaling the generation rate by the dimension.
\newline

The simulation does not take into account the common dead time. The validity of the security proof assumes that all detectors have efficiency 100\% at the moment of detection. The dead time simulated in this work only assumes the dead time for one detector. This is a vulnerability that the eavesdropper can take advantage of. To avoid this, after a detection event, all detectors should be shut down for the duration of the dead time, or the detections that occur during this time should be discarded. This, in fact, would make the gap between consecutive dimensions more advantageous for high dimensions, as the number of discarded detections for them would be much smaller. Considering this, the simulation represents the worst case scenario. 
\\
The secret key length is calculated from the finite-key approach, where the length of the secret key is equal to:

\begin{eqnarray}
l\leq log_2(d)\cdot s_{Z,0}^{l} +  s_{Z,1}^{l}\cdot(c - H_d(\phi_Z)) - \Delta_{EC} \nonumber \\ 
- a \cdot log_2\left (\frac{b}{\epsilon_{sec}} \right) - log_2\left (\frac{2}{\epsilon_{cor}}\right)
\end{eqnarray}

\noindent
where $d$ is the number of dimensions, $s_{Z,0}^{l}$ and $s_{Z,1}^{l}$ are the lower bounds for the zero-photon and single-photon events in the Z basis, respectively. $H_d()$ is the Shannon entropy function for the number of dimensions $d$ \cite{efficient_EC}:
\begin{equation}
    H_d(x) = -x\cdot log_2\left(\frac{x}{d-1}\right) - (1 - x)\cdot log_2(1 - x),
\end{equation}
$\phi_Z$ is the phase error rate, $a$ and $b$ are the decoy state parameters (for one decoy state protocol $a = 6$ and $b = 19$, $\Delta_{EC}$ is the information lost due to error correction, and $\epsilon_{sec}, \epsilon_{cor}$ are the secrecy and correctness parameters \cite{origfinitekey}, which we chose to be $10^{-9}$ and $10^{-15}$.\\
For numerical calculation, we assume that $H(0)=0$ \cite{quantumcommunication_nielsen}. The secret key length is calculated according to ref. \cite{1decoy} and \cite{Efficient} and is based on the channel attenuation $\eta$. In all subsequent plots we express channel loss in decibel units, i.e. $L_{dB}=-10log_{10}(\eta)$.

An additional loss factor $\eta_p(d)$ arises due to post-processing of interference peaks in the passive interferometric state analysis. This implementation-specific loss factor $\eta_p(d)=0.5^{k_j}$, where $k$ is the number of cascading interferometers in basis j, reduces the secure key rate of higher-dimensional implementations at zero channel loss.  Note that this could be mitigated by adopting a post-selection-free detection approach \cite{Vedovato.2018} based on fast electro-optic switching in the high-dimensional superposition basis.  

The probability $\tau_n$ of generating a state with mean photon number $\mu_i$, and the photon number n is:
\begin{equation}
    \tau_n = \sum_i\frac{p_{\mu_i} e^{-\mu_i}\mu_i^n}{n!}
\end{equation}
where $\mu_i$ is the mean photon number for the signal or decoy state $i$.
The lower bound on the single-photon events in the Z basis:
\begin{equation}
\begin{aligned}
    & s_{Z,1}\ge s_{Z,1}^l:=\\
    & \frac{\tau_1 \mu_1}{\mu_2(\mu_1 - \mu_2)} 
    \left(n_{Z,\mu_2}^- - \frac{\mu_2^2}{\mu_1^2}n_{Z,\mu_1}^+ \nonumber 
    - \left(\frac{\mu_1^2 - \mu_2^2}{\mu_1^2} \right)\frac{s_{Z,0}}{\tau_0} \right)
\end{aligned}
\end{equation}
The fluctuations of the number of events in the Z basis from Hoeffding's inequality ($i\subset \mu_1, \mu_2$):
\begin{equation}
    n_{Z,i}^{\pm}:=\frac{e^i}{p_i}\left (n_{Z,i} \pm \sqrt{\frac{n_Z}{2}log\frac{1}{\epsilon_1}} \right)
\end{equation}
where $p_i$ is the probability of choosing a state with a mean photon number $\mu_1$.
The number of errors:
\begin{equation}
    m_{Z,i}^{\pm}:=\frac{e^i}{p_i} \left (m_{Z,i} \pm \sqrt{\frac{m_Z}{2}log\frac{1}{\epsilon_2}}\right)
\end{equation}
The upper bound on the zero-photon events for dimensionality d is:
\begin{equation}
    \begin{aligned}
    & s_{Z,0} \leq s_{Z,0}^u:= \\
    & \frac{1}{\left(1-\frac{1}{d}\right)}\tau_0 \frac{e^{\mu_2}}{p_{\mu_2}}\left(m_{Z,\mu_2} +\sqrt{\frac{m_Z}{2}log\frac{1}{\epsilon_2}} \nonumber
    + \sqrt{\frac{n_Z}{2}log \frac{1}{\epsilon_1}}\right)
    \end{aligned}
\end{equation}
The lower bound on the zero-photon events:
\begin{equation}
    s_{Z,0} \ge s_{Z,0}^l:=
    \frac{\tau_0}{\mu_1 - \mu_2}\left(\mu_1 n_{Z,\mu_2}^{-} - \mu_2 n_{Z,\mu_1}^{+}\right)
\end{equation}
The probability of detecting a state with a mean photon number $\mu_i$:
\begin{equation}
    P_{Z,det,\mu_i}:= c_{DT}P_Z p_{\mu_i}\left(\left(1 - e^{-\mu_i \eta}\right) + P_{DC}\right)
\end{equation}
where $c_{DT}$ is the dead time coefficient, $P_Z$ is the probability to choose the Z basis, $P_{DC}$ is the probability of a dead count. $P_{Z,det,tot}$ is the sum of all the probabilities of detecting a state with a mean photon number $\mu_i$.
The dead time coefficient:
\begin{equation}
    c_{DT} := \frac{1}{1 + RP_{Z,det,tot}t_{DT}}
\end{equation}
The equations above form an iterative relation. In the optimization, only one round was calculated, which is consistent with the results from ref. \cite{1decoy}.\\
The probability of getting an error for a state with a mean photon number $\mu_i$:
\begin{equation}
\begin{aligned}
    & P_{Z,err,\mu_i}:= \\
    & c_{DT}P_Z P{\mu_i}\left(\left(1 - e^{-\mu_i \eta}\right) + P_{err} + P_{DC}\left(1-\frac{1}{d}\right) \right)
\end{aligned}
\end{equation}

\noindent
where $P_{err}$ is the probability of an error (1\%).
$P_{Z,err,tot}$ is the sum of all the probabilities of detecting a state with a mean photon number $\mu_i$.
Number of detection events and errors for each mean photon number $\mu_i$:
\begin{equation}
    n_{Z,\mu_i}:=n_Z \cdot \frac{P_{z,det,\mu_i}}{P_{Z,det,tot}}
\end{equation}

\begin{equation}
    m_{Z,\mu_i}:=n_Z \cdot \frac{P_{z,err,\mu_i}}{P_{Z,det,tot}}
\end{equation}

Phase error rate: 
\begin{equation}
    \phi_Z \leq \phi_X^u:= \frac{v_{X,1}^u}{s_{X,1}^l} + \gamma \left(\sqrt{\epsilon_{1}},s_{Z,1}^l,s_{X,1}^l \right)
\end{equation}

\begin{equation}
    \gamma(a,b,c) :=\sqrt{\frac{b + c}{bc}\frac{1+c}{c}log\frac{1}{a}}
\end{equation}
Number of errors in the X basis:
\begin{equation}
    v_{X,1} \leq v_{X,1}^u := \frac{\tau_1}{\mu_1 - \mu_2}\left(m_{X,\mu_1}^+ - m_{X,\mu_2^-}\right)
\end{equation}
For this calculation, the parameters in the X basis are determined in the same way as the Z basis parameters. Due to the symmetrical character of the Shannon entropy function, $\phi_Z$ is bound in the following way:

\begin{center}
\begin{tabular}{||c c c||} 
 \hline
 D & x & $\phi_Z$ \\ [0.5ex] 
 \hline\hline
 2 & 0.5 & 1.0 \\ 
 \hline
 4 & 0.75 & 2.0 \\
 \hline
 8 & 0.87 & 3.0 \\
 \hline
 16 & 0.95 & 4.0 \\
 \hline
\end{tabular}
\end{center}
where D is the number of dimensions and x is the input for $\phi_Z$. Above the x input value, the output of the $\phi_Z$ function should be bounded to the listed value.

\begin{equation}
    n_X := n_Z \frac{P_X}{P_Z}
\end{equation}
\begin{equation}
    \frac{m_X}{n_Z} := \frac{P_{Z,err,tot}}{P_{Z,det,tot}}
\end{equation}
The estimated information leakage:
\begin{equation}
    \Delta_{EC} := 1.08\cdot n_Z\cdot H_d \left(\frac{P_{Z,err,tot}}{P_{Z,det,tot}} \right)
\end{equation}

\begin{equation}
    N_{tot} := \frac{n_Z}{P_{Z,det,tot}}
\end{equation}

\begin{equation}
    SKR := \frac{l}{N_{tot}}R
\end{equation}

error tolerance values \cite{origfinitekey}:
\begin{eqnarray}
    \epsilon_1 = \epsilon_{sec}/19 \nonumber \\
    \epsilon_2 = \epsilon_{cor}/19
\end{eqnarray}

The following parameters were bound to 0 if the result of their calculation was negative: $n_{Z,i}^{-}, m_{Z,i}^{-}, s_{Z,0}^l, s_{Z,1}^l$.

\bibliography{Ref}

\end{document}